\documentclass{elsart}
\usepackage{amssymb}
\usepackage{graphicx}

\def\E#1{\hbox{$10^{#1}$}}
\def\mic  {\hbox{$\mu$m}}
\def\deg    {\hbox{$^\circ$}}
\def\about  {\hbox{$\sim$}}
\def\sub#1{_{\rm #1}}
\def\tV     {\hbox{$\tau\sub V$}}
\def\Ri     {\hbox{$R\sub{i}$}}
\def\Mo     {\hbox{$M_\odot$}}
\def\Lo     {\hbox{$L_\odot$}}
\def\ga     {\hbox{$\gtrsim$}}
\def\la     {\hbox{$\lesssim$}}
\begin{document}

\begin{frontmatter}

\title{IR Emission from AGNs}

\author{Moshe Elitzur\thanksref{LAOG}}
\address{Physics \& Astronomy Dept., University of Kentucky,
         Lexington, KY 40506, USA,
         and LAOG,  BP 53, F-38041 Grenoble, France}
\ead{moshe@pa.uky.edu}

\thanks[LAOG]{
On sabbatical leave at the Laboratoire d'Astrophysique Observatoire de
Grenoble.
}

\begin{abstract}
Unified schemes of active galactic nuclei (AGN) require an obscuring dusty
torus around the central engine. Torus sizes of hundreds of parsecs were
deduced from early theoretical modeling efforts, but high-resolution IR
observations now show that the torus size is no more than a few parsecs. This
conflict is resolved when the clumpy nature of the torus is taken into account.
The compact torus may be best understood when identified with the dusty,
optically thick region of the wind coming off the central accretion disk.
\end{abstract}


\end{frontmatter}

\section{Introduction}
 \label{sec:introduction}

The great diversity of AGN classes has been explained by a single unified
scheme (e.g.\ Antonucci 1993; Urry \& Padovani 1995). The nuclear activity is
powered by a super\-massive (\about\E6--\E{10} \Mo) black hole and its
accretion disk. This central engine is surrounded by dusty clouds, which are
individually optically thick, in a toroidal structure (Krolik \& Begelman
1988). Much of the observed diversity is simply the result of viewing this
axi\-symmetric geometry from different angles. The clumpy torus provides
anisotropic obscuration of the central region so that sources viewed face-on
are recognized as type 1 objects, those observed edge-on are type 2. The
fraction of the sky obscured by the torus determines the relative numbers of
type 1 and 2 sources. From the statistics of Seyfert galaxies, Schmitt et al
(2001) find that the torus height and radius obey $H/R$ \about\ 1. In the
ubiquitous sketch by Urry \& Padovani (1995), the torus is depicted as a large
doughnut-like object, presumably populated by molecular clouds accreted from
the galaxy. Gravity controls the orbital motions of the clouds, but the origin
of cloud vertical motions capable of sustaining the ``doughnut" in a
steady-state with $H \sim R$ was recognized as a problem by Krolik \& Begelman
(1988). This problem has eluded solution to this date.

\section{The Torus Size}

Obscuration does not depend individually on either $H$ or $R$, only on their
ratio. To determine an actual size one must rely on the torus emission. In the
absence of high-resolution IR observations, early estimates of the torus size
came from theoretical analysis of the spectral energy distribution (SED). Pier
\& Krolik (1992) preformed the first detailed calculations of dust radiative
transfer in a toroidal geometry. Because of the difficulties in modeling a
clumpy medium, Pier \& Krolik approximated the density distribution with a
uniform one instead. They concluded that the torus has an outer radius $R$
\about\ 5--10 pc, but later speculated that this compact structure might be
embedded in a much larger, and more diffuse, torus with $R$ \about\ 30--100 pc
(Pier \& Krolik 1993). Granato and Danese (1994) extended the smooth-density
calculations to more elaborate toroidal geometries.  From comparisons of their
model predictions with the observed IR emission at $\lambda$ \about\ 10--25
\mic\ they concluded that the torus must have an outer radius $R$ \ga\
300--1000 pc, and that its radial density profile must be constant; later,
Granato et al (1997) settled on hundreds of pc as their estimate for the torus
size. Subsequently, $R >$ 100 pc became common lore.

The advent of high-resolution IR observations brought unambiguous evidence in
support of Pier \& Krolik's original proposal of compact torus dimensions. The
best studied source is NGC1068. K-band imaging shows that the emission region
has $R\ \la$ 1 pc (Weigelt et al 2004). VLTI interferometry shows that the 10
\mic\ flux comes from a hot ($T >$ 800 K) central region of $\sim 1$ pc and its
cooler ($T \sim$ 320 K) surrounding within $R \sim$ 2--3 pc (Jaffe et al 2004).
Keck observations established that $R < $ 17 pc at 8--25 \mic\ (Bock et al
2000). Recent Gemini observations by Mason et al (2005) confirm that $R < $ 15
pc at 12\mic\ and show that mid-IR observations at larger apertures are
dominated by extended, low-brightness dust emission from the ionization cones.
Similarly compact dimensions are found in high-resolution IR observations of
other AGN: Circinus --- $R$ \about\ 1 pc at 2.2 \mic\ (Prieto et al 2004) and
$R$ \la\ 2 pc at 8.7 \mic\ and 18 \mic\ (Packham et al 2005); NGC4151
--- Swain et al (2003) find marginally resolved 2.2 \mic\ emission with $R \le$
0.05 pc, and Radomski et al (2003) find $R < $ 17 pc at 10 \mic\ and 18 \mic;
NGC1097 --- $R <$ 5 pc in near IR (Prieto et al 2005).

%%%%%%%%%%%%%%%%%%%% T--R %%%%%%%%%%%%%%%%%%%%%%%%%%%%%%%%
\begin{figure}[ht]
\centering \leavevmode
 \includegraphics[width=0.7\hsize,clip]{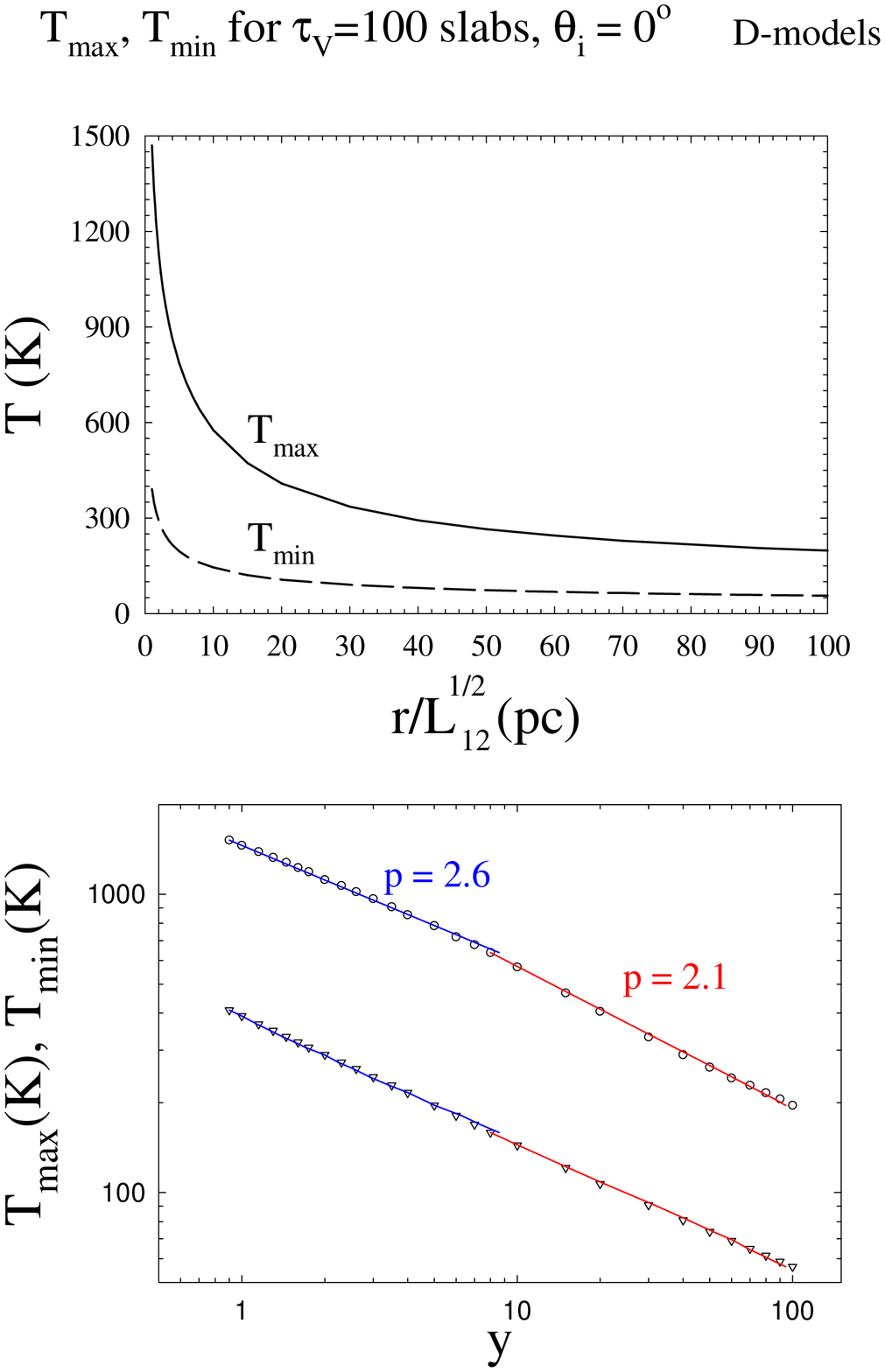}
\caption{The highest ($T_{\rm max}$) and lowest ($T_{\rm min}$) dust
temperatures on the surface of an optically thick cloud at distance $r$ from an
AGN with luminosity $L_{12} = L/\E{12}\Lo$. The highest temperature occurs on
the illuminated face, the lowest on the dark side (from Nenkova et al 2005).
\label{fig:T-r}}
\end{figure}
%%%%%%%%%%%%%%%%%%%%%%%%%%%%%%%%%%%%%%%%%%%%%%%%%%%%%%%%%%%%%%

It can be argued that IR observations only determine the size of the
corresponding emission region and that the actual torus size could in fact be
much larger, but mid-IR flux considerations were the sole reason for
introducing large sizes in the first place. In addition to IR, detailed studies
of the central regions of active galaxies were made in CO observations.
Schinnerer et al (2000) trace rotating molecular clouds in NGC1068 down to a
distance of about 13 pc from the nucleus. From the velocity dispersions they
find that at $R \simeq$ 70 pc, the height of the molecular cloud distribution
is only $H$ \about\ 9--10 pc, for $H/R$ \about\ 0.15. Thus, although resembling
the putative torus, the distribution of these clouds does not meet the
unification scheme requirement $H/R$ \about\ 1; evidently, these clouds reside
outside the torus, in accord with the IR observations.

In establishing compact torus sizes, IR observations have eliminated the only
rationale for large dimensions and uncovered a fundamental discrepancy with the
modeling results. Dust emission at 10 \mic\ requires $T$ \about\ 200--300 K, in
turn implying large distances from the heating source. The constant radial
density profiles and large dimensions deduced from the torus modeling merely
reflected the large amounts of cool dust needed to produce the observed IR
flux. Since dust temperature is uniquely related to distance from the AGN in
smooth density distributions, it is impossible to get around this fundamental
difficulty. However, in clumpy media the one-to-one correspondence between
distance and temperature no longer holds. The temperature of an optically thick
dusty cloud is much higher on the side illuminated by the AGN than on the dark
side (Nenkova et al 2002). In contrast with smooth density distributions, in a
clumpy medium with cloud sizes much smaller than radial distances,

\begin{itemize}
\item
Different dust temperatures coexist at the same distance from the AGN

\item
The same temperature occurs at different distances --- the dark side of a cloud
close to the AGN can be as warm as the bright side of a farther cloud
\end{itemize}

%%%%%%%%%%%%%%%%%%%%%%%%%%%%%%%%%%%%%%%%%%%%%%%%%%%%%%
\begin{figure}
\centering \leavevmode
 \includegraphics[width=0.7\hsize,clip]{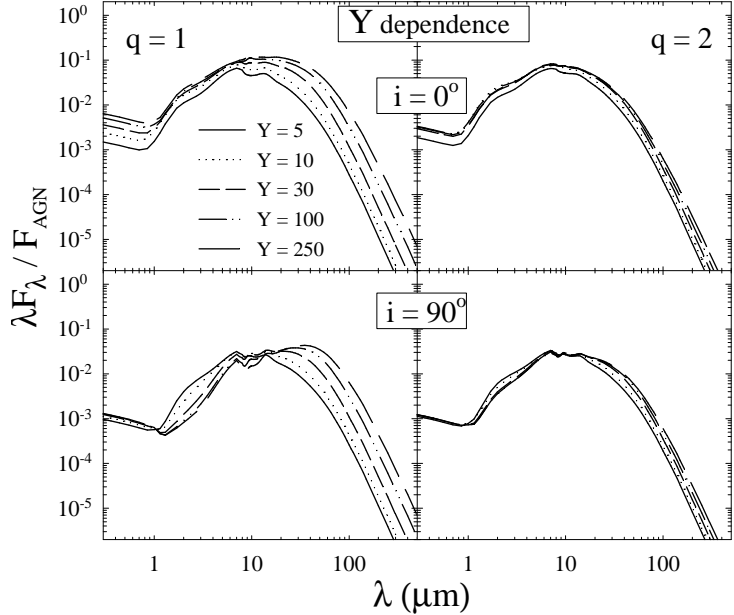}
\caption{Dependence of the SED on the torus radial thickness $Y = R/\Ri$, where
the inner radius \Ri\ is determined by dust sublimation at 1500\,K. The torus
is comprised of clouds, each with visual optical depth \tV\ = 60, and has a
total of 5 clouds, on average, along radial equatorial rays. The clouds radial
distribution is $\propto 1/r^q$, with $q$ = 1 and 2 as indicated, and their
angular distribution is Gaussian with $\sigma = 45$\deg. Pole-on viewing is $i$
= 0\deg, edge-on $i$ = 90\deg. The torus size has a negligible effect in the
case of the steeper power law $q = 2$ (from Nenkova et al 2005).}
\label{Fig:SED-Ydep}
\end{figure}
%%%%%%%%%%%%%%%%%%%%%%%%%%%%%%%%%%%%%%%%%%%%%%%%%%%%%%

Figure \ref{fig:T-r} shows the distance variation of temperature on the bright
and dark sides of an optically thick cloud.  When the illuminated face is at
$T$ = 900 K, the dust temperature on the dark side is only $T$ \la\ 300 K.
Therefore, even at the compact dimensions established by the observations, AGN
tori contain cool dust thanks to their clumpy nature. Detailed modeling of IR
emission from clumpy tori (Nenkova et al 2002, 2005) easily produce SED similar
to those observed in AGN even for $R$ as small as 5 pc. Figure
\ref{Fig:SED-Ydep} shows that varying the radial size of a clumpy torus by as
much as a factor of 50 has only a small impact on its SED. The effect of
overall size is confined mostly to long wavelengths when the radial cloud
distribution varies as $1/r$, disappearing altogether for the steeper
distribution $1/r^2$.

\section{The Torus as a Disk Wind}

Compact torus sizes give strong impetus for an alternative scenario to the
steady-state ``doughnut", involving cloud outflow in hydromagnetic disk winds.
In this approach, which was advanced by numerous authors (Emmering et al 1992;
K\"onigl \& Kartje 1994; Bottorff et al 1997, 2000; Kartje at al 1999), the
proverbial torus is merely a particular region in the wind which happens to
provide the toroidal obscuration required by unification schemes. The two
scenarios differ fundamentally. In the first, clouds flowing in from the galaxy
settle into closed-orbit motions around the center, their velocities containing
comparable vertical and rotational components.  The result is a puffed-up
steady-state structure around the central engine --- the ``doughnut". It is a
well-defined component of the system, separate from the accretion disk and any
wind that might be coming off its surface. In contrast, in the torus-as-a-wind
scenario the torus is simply that region of the clumpy wind wherein the clouds
are dusty and optically thick. Cloud inflow from the galaxy does feed the
central accretion disk, but it is not presumed to produce a separate puffed-up
structure, which need not exist at all in this approach.

The disk-wind scenario does not suffer from the vertical-support problem of the
``doughnut", but questions involving the cloud-uplift and wind-driving
mechanisms are still unsettled. However, although the theoretical issues are
not yet fully resolved, observations give ample evidence for winds and cloud
motions in AGN (see, e.g., Elvis 2004), and even provide support for a
disk-wind geometry (Hall et al 2003). In particular, recent H$_2$O maser
observations in NGC3079 provide evidence for the uplift of molecular clouds off
the disk surface and their subsequent outflow along rotating streamlines
(Kondratko et al 2005). We are currently attempting to incorporate in the disk
wind scenario the observational constraints on the clumpy obscuration (Elitzur
\& Shlosman, in preparation). If successful, the outcome will be a ``grand
unification scheme" in which the clouds responsible for disparate phenomena
such as broad emission and absorption lines, warm absorption, and the toroidal
obscuration are all members of a single clumpy disk wind, distinguished from
each other only by their physical properties which are controlled by location
in the outflow.

\ack The author's research is supported by NASA and NSF. The warm hospitality
at LAOG is gratefully acknowledged.

\end{document}